\documentclass[%
aip,
citeautoscript,
 amsmath,amssymb,
reprint,%
floatfix]{revtex4-1}
\usepackage{graphicx}
\usepackage{dcolumn}
\usepackage{bm}
\usepackage[utf8]{inputenc}
\usepackage[T1]{fontenc}
\usepackage{mathptmx}
\usepackage{etoolbox}

\usepackage{orcidlink} 
\usepackage[letterpaper, margin=1in]{geometry}

\usepackage[T1]{fontenc}
\usepackage[utf8]{inputenc}
\usepackage{lmodern}
\usepackage{microtype}

\usepackage{amsmath}
\usepackage{amsfonts}
\usepackage{amssymb}
\usepackage{nicefrac}
\usepackage{upgreek}
\usepackage{array}
\usepackage{tabularx}

\usepackage{booktabs}
\usepackage{caption}
\usepackage{float}
\usepackage{xcolor}
\usepackage{colortbl}
\usepackage{graphicx}
\usepackage{subcaption}
\usepackage{multirow}

\usepackage[english]{babel}
\usepackage{url}
\usepackage{hyperref}
\hypersetup{
    colorlinks=true,
    linkcolor=blue,
    citecolor=blue,
    urlcolor=blue,
}

\begin{document}

\preprint{AIP/123-QED}

\title{A reaction volume bias Monte Carlo trial for sampling chemisorption in confinement}

\author{Samiha Sharlin\,\orcidlink{0000-0002-6379-9206}}
\thanks{These authors contributed equally to this work.}
\affiliation{Department of Chemical, Biochemical, and Environmental Engineering,\\ University of Maryland Baltimore County, Baltimore, MD 21250}
\author{Harold W. Hatch\,\orcidlink{0000-0003-2926-9145}}
\thanks{These authors contributed equally to this work.}
\affiliation{Theory, Modeling, and Simulation Group, Material Data Division,\\ National Institute of Standards and Technology, Gaithersburg, MD 20899-8380}
\author{Daniel W. Siderius\,\orcidlink{0000-0002-6260-7727}}
\affiliation{Theory, Modeling, and Simulation Group, Material Data Division,\\ National Institute of Standards and Technology, Gaithersburg, MD 20899-8380}
\author{Tyler R. Josephson\,\orcidlink{0000-0002-0100-0227}}
\email{tjo@umbc.edu}
\affiliation{Department of Chemical, Biochemical, and Environmental Engineering,\\ University of Maryland Baltimore County, Baltimore, MD 21250}
\affiliation{Department of Computer Science and Electrical Engineering,\\ University of Maryland Baltimore County, Baltimore, MD 21250}

\date{\today}

\begin{abstract}
\textcolor{red}{(This document has not been peer reviewed but has been cleared by NIST for release.)}
Molecular modeling of chemisorption with Monte Carlo requires the development of new trial moves to efficiently sample complex fluids such as water in Brønsted acid zeolites.
Here, we develop a reaction volume bias (RxVB) Monte Carlo trial for modeling chemisorption by combining identity-switch and aggregation-volume-bias (AVB) moves.
This method aims to promote the sampling of reactions by choosing reactive pairs that are within an arbitrarily specified reaction volume.
The RxVB move achieves up to a 90-fold increase in accepted reaction events over unbiased moves in a single-site slit-pore model, corresponding to a 70-fold gain in statistical efficiency after accounting for computational overhead.
But when the same move is applied to water in a Brønsted acid zeolite without orientational bias, there is no measurable speedup for a single MFI unit cell.
We demonstrate a simple expression that predicts the maximum efficiency increase in the simplest case where selecting reactants that are near each other is the major sampling bottleneck.
Dense water systems may require additional configuration-bias or orientational bias to improve sampling of the hydrogen bond network in order to increase acceptance.
This new RxVB trial was made available with examples in the open-source Free Energy and Advanced Sampling Simulation Toolkit (FEASST) simulation package.

\begin{figure}[H]
\begin{centering}
\includegraphics[width=0.7\textwidth]{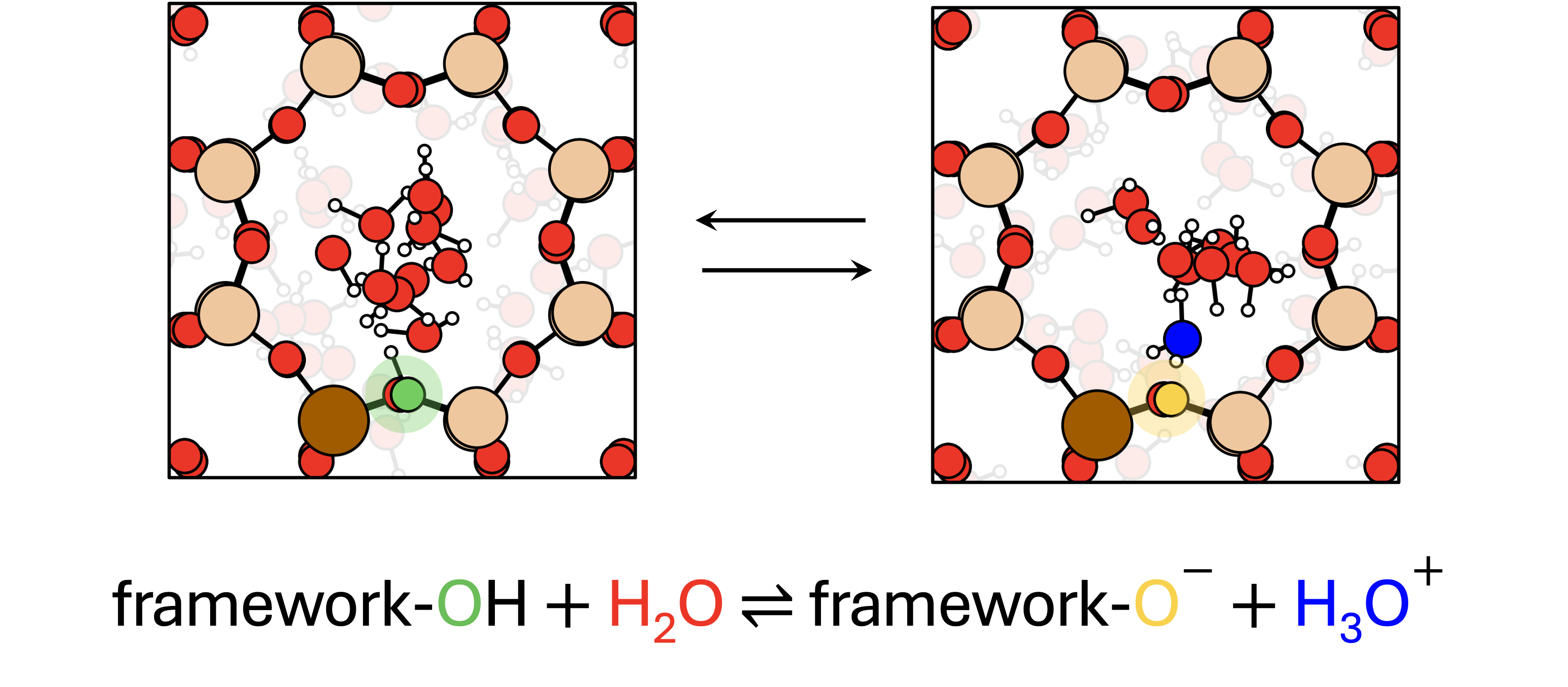}
\caption*{Table of Contents Graphic}
\label{toc}
\end{centering}
\end{figure}
\end{abstract}

\maketitle

\vspace{1em}
\section{Introduction}

Liquid-phase catalysis is affected by the solvation of active sites, as well as the reactants and products involved. 
In nanoporous materials like zeolites, metal-organic frameworks (MOFs), and activated carbon, confinement drastically changes the solvation behavior compared to bulk conditions \cite{bhushan1995nanotribology, schoen1998analytical, stewart2012phase}. 
For instance, nanoporous materials reorganize the structure of water and modify local activities near Br{\o}nsted acid sites, creating non-bulk hydrogen-bond networks such as Eigen, Zundel, and chainlike motifs \cite{munoz2013nanoconfinement, lynch2020water, leoni2021nanoconfined, stanciakova2021water,liu2024water}. Additionally, confinement also reduces the effective dielectric constant and enhances ion pairing, which influences the stabilization of charged intermediates \cite{hennequin2021competition,aluru2023fluids, wang2019genesis, gao2024reorientation}. 
These combined effects shift the apparent acidity and alter both reaction equilibria and activation barriers \cite{grifoni2021confinement,van2023operando}. 

Accurately quantifying solvation effects poses significant challenges, as even binary adsorption experiments result in under-determined mass and volume balances to quantify both solute \emph{and} solvent co-adsorption \cite{dejacoVaporLiquidPhase2020}. While advanced characterization techniques can locate molecules inside pores, they are costly, difficult, and only applicable for select adsorbate/adsorbent combinations \cite{nairLocationMxyleneSilicalite2000, petersen2022review, rowsellGasAdsorptionSites2005, vaidhyanathanDirectObservationQuantification2010,baekHightemperatureSituCrystallographic2015}. 
Ab initio molecular dynamics (AIMD) simulations allow modeling of these interactions, but their high computational cost limits simulations to short durations and small systems, and also requires a prior assumption of the number and type of solvent molecules present \cite{vanspeybroeckRecentDevelopmentComputational2003, stanciakovaCooperativeRoleWater2019, bukowskiDefectMediatedOrdering2019, batesStructureSolvationConfined2020, reyDynamicFeaturesTransition2020}. 

In contrast, classical Monte Carlo (MC) methods enable simulations where molecules can enter and exit adsorption sites at equilibrium, providing useful data such as adsorption isotherms and free energies of transfer at reaction temperatures and pressures \cite{smit2008molecular, xiong2012molecular, dejacoAdsorptiveSeparation1Butanol2016, 
yang2017separation, 
bereciartua2017control, josephsonAdsorptionFuranHexanoic2021}.
These molecular insights from modeling are challenging, if not impossible, to obtain experimentally. Nonetheless, MC simulations are predominantly conducted on porous materials without catalytically active sites, as interactions involving chemisorption are not easily sampled, and they are also not fully characterized with existing force fields.

First-principles MC simulations can predict chemisorption in principle, but each individual simulation is expensive enough that only a handful of configurations can be sampled on exascale machines \cite{fetisovUnderstandingReactiveAdsorption2018, fetisovFirstPrinciplesMonte2018, baiFirstPrinciplesGrandCanonicalSimulations2021}.
Efficiently sampling reactive events remains a bottleneck for liquid-phase heterogeneous catalysis, where reactants must first find each other in confined space before they can exchange identities. 
From proton shuttling in fuel-cell membranes to ion pairing in electrolyte solutions, brute-force MC sampling spends most of its time on trials involving non-reactive configurations \cite{torrie1977nonphysical, gao1994simulation, voth2006computer}. 
Development of strategic MC moves enables sampling of phenomena not previously possible, as they can enable orders-of-magnitude improvements in sampling efficiency at minimal cost.
For example, Configurational Bias Monte Carlo (CBMC) moves allow for efficient sampling of long, flexible chain molecules by inserting the chain incrementally \cite{siepmannConfigurationalBiasMonte1992, vlugtImprovingEfficiencyConfigurationalbias1998}. 
This approach favors energetically favorable configurations, minimizing overlaps with the framework and other particles.

Prior work with MC on predicting adsorption in Br{\o}nsted acid zeolites has largely used unreactive or coarse-grained acid-site representations that fit other adsorbates but do not account for water interactions \cite{hunger2002characterization,calero2006coarse,janda2016effects}, which is one of the most important adsorbates. 
To accurately simulate water chemisorption in these frameworks, a model should account for proton transfer and the hydrogen bond networks \cite{datar2021monte, siderius_flat-histogram_2024}. 

In this article, we develop a reaction volume biased (RxVB) Monte Carlo move, which biases the selection of nearby pairs of reactants as described in Section~\ref{sec:rxnmc} and we implemented RxVB in the open-source Free Energy and Advanced Sampling Simulation Toolkit (FEASST) software package \cite{hatch_monte_2024}.
The RxVB move is demonstrated in Section~\ref{slitpore_model} using a simple fixed reaction site in a slit pore with a single-site square-well fluid in order to validate the efficiency metrics and develop an expected upper bound to the possible gains in efficiency, which exceed 90-fold in some cases.
We then apply the RxVB move to water in a Mobile Five (MFI) zeolite with a single aluminum-substituted Br{\o}nsted acidic site in Section~\ref{sec:water}, but increase sampling efficiency was not observed when the hydrogens were randomly oriented about their bonded oxygen during the reaction due to the constraint of rigid bond lengths and angles.
Finally, we conclude our findings in Section~\ref{sec:conclusion}.
\section{Reactive Monte Carlo\label{sec:rxnmc}}

Traditional reactive Monte Carlo (RxMC) methods use MC moves to switch predefined molecular species and achieve equilibrium between reactants and products \cite{johnson1994reactive, smith1994reaction, glotzer1994monte, hansen2005reactive, heath2008simulation}. 
To address sampling challenges with insertion and deletion, earlier studies used CBMC to build whole molecules in one step \cite{consta1999recoil, jakobtorweihen2006combining, bai2017assessment}, or Continuous Fractional Component (CFC) MC to grow them gradually \cite{shi2007continuous,shi2008improvement,balaji2015simulating,torres2017behavior,poursaeidesfahani2017efficient}. This works well in bulk, but in confinement, efficiency may be increased if MC moves where placed molecules in a way that reflects the locality of chemisorption.

A RxVB move can be more effective for sampling chemisorption events when reactants and products have strong energetic interactions or are entropically favored to react in close proximity.
In a traditional MC identity switch, particles are selected from anywhere and each is deleted and regrown in the location of the other \cite{de1989phase, panagiotopoulos1989exact, panagiotopoulos1995gibbs, martin1997predicting}. 
We hypothesize that a more efficient sampling is possible by performing a ``local identity switch'' between reactants and products inside a predefined reaction shell.

The earliest works in biased MC sampling methods that preferentially move molecules into or out of target-specific regions include bond-bias (BB) MC \cite{tsangaris1994bond}, association-bias (AB) MC \cite{busch1996monte}, monomer addition subtraction algorithm (MASA) \cite{visco1999modeling}, unbound–bound (UB) MC \cite{wierzchowski_general-purpose_2001, wierzchowski_ub_2002} and Aggregation-Volume-Bias MC (AVBMC) \cite{chen2000avbmc, chen2001avbmc2}.
The AVBMC method is simple and transferrable with variations reported in literature such as AVBMC4,\cite{siderius_flat-histogram_2024, hatch2025best} which only samples between two cluster regions that may also be overlapping, and orientational specific AVB for patchy particles that have a repulsive core with short-range directional interaction sites \cite{rovigatti_how_2018}.
Integrating geometric pre-selection (AVBMC) with a standard identity switch has not previously been applied to acid–base chemisorption in MC simulations to our knowledge.
The acid–base interactions that have been investigated with constant-pH methods in MC treat the proton as an implicit reservoir and flip HA\,$\rightleftharpoons$\,A$^-$ with an acceptance based on $\mathrm{pH}-\mathrm{p}K_a$ \cite{reed1992monte, landsgesell2017simulation}. These schemes are efficient for titration curves without explicit ions, but as they omit explicit hydronium/co-ions, they limit the study of confinement effects and hydration near Br{\o}nsted sites. 
Another scheme, the Grand-Reaction Method (GRM), instead couples the simulation box to a pH/salt reservoir by combining reaction moves with grand-canonical insertion/deletion of electroneutral ion pairs (and water) \cite{landsgesell2020grand, beyer2023generalized}. 
Pairing GRM with RxVB could simultaneously sample local chemisorption near sites while maintaining reservoir composition, enabling direct calculation of partitioning coefficients and pore-specific titration curves.

Another related approach, explicit-bond RxMC, uses local particle selection (e.g., an inclusion sphere) to propose topology changes by creating or deleting a real bond between two sites \cite{blanco2024explicit}. 
The Metropolis acceptance for explicit-bond RxMC is written so that the equilibrium constant accounts for bond formation, while the bond’s explicit potential energy is excluded from $\Delta U$ to avoid double counting. 
Explicit-bond RxMC is ideal when connectivity itself is the observable, for example, dimerization, cross-linking, or polymer growth. 
For observing protonation states and hydration in water chemisorption with active sites, identity-based sampling is more straightforward.

In the standard formulation of RxMC, the Metropolis acceptance for a reaction A$+$B $\to$ C$+$D requires the individual chemical potentials $\mu_i$ for every reactant and product, which are typically constructed from ideal-gas partition functions \cite{mcquarrie1997physical, hansen2005reactive} or JANAF tables \cite{chase1985janaf}. However, insufficient experimental data often makes this process complicated.
In contrast, first-principles RxMC models the system as independent atoms that combine into molecules and avoids partition function tables, but its computational cost approaches that of AIMD, limiting sampling even on large machines \cite{fetisovUnderstandingReactiveAdsorption2018, fetisovFirstPrinciplesMonte2018, baiFirstPrinciplesGrandCanonicalSimulations2021}.
Reactive MD with bond-order potentials (e.g., ReaxFF) and first-principles MD can also be used, but controlling chemical potentials and reaching equilibrium mixtures in confinement is also computationally expensive.

The Metropolis acceptance of RxMC for a predefined reaction from $A{+}B\to C{+}D$ is written in a semigrand, identity-switch form and is controlled by a single parameter, the reaction chemical potential difference, $\Delta\mu_{\mathrm{rxn}}$, which is related to the standard-state equilibrium constant, $K^\circ(T)$:
\begin{equation}\label{rxmc_eqn}
  \Delta\mu_{\mathrm{rxn}}
  \;\equiv\;
  \mu_C + \mu_D - \mu_A - \mu_B
  \;=\;
  -\,k_{\mathrm{B}}T \ln K^\circ(T),
\end{equation}
where $T$ is the temperature and $k_B$ is the Boltzmann constant.
We can either set $\Delta\mu_{\mathrm{rxn}}$ from experiment, \emph{ab initio} free energies, or study dependence of the reactant and product concentrations on $\Delta\mu_{\mathrm{rxn}}$ to match target concentrations.

\subsection{Acceptance of unbiased reaction and reaction volume bias Monte Carlo\label{sec:acceptance}}

In this article, we derive and implement a novel Monte Carlo reaction trial which biases the reactants or products to be within a predefined local region (RxVB).
But first, the more simple unbiased reaction of A + B $\rightleftharpoons$ C + D proceeds as follows, and was considered in order to quantify the speedup of the RxVB move.
First, the forward or reverse reaction trial is randomly chosen.
In the forward reaction, a random particle of type A is chosen from among the $N_A$ particles.
Then a random particle of type B is chosen from among the $N_B$ particles.
Reject the move if $N_A = 0$ or $N_B = 0$.
Particle A then transforms into C.
For single site particles, this transformation puts the center of the new C particle at the center of the former A particle.
For a rigid multiple site particle (e.g., framework-OH or water), the position of a predefined site (e.g., oxygen) of C (e.g., framework-O$^-$ or hydronium) is placed in the former position of a predefined site of the A particle (e.g., oxygen), and then the C particle is randomly rotated about that site.
This assumes all particles are rigid and have no intramolecular degrees of freedom.
Particle B transforms into D in a similar fashion.
The reverse trial selects C and D to transform into A and B in a similar fashion with random orientations.
Application of local detailed balance\cite{hatch2025best} to the Metropolis Monte Carlo acceptance,\cite{metropolis_equation_1953} $\chi$, in the semi-grand canonical ensemble yields the following for the forward and reverse trials, respectively,
\begin{equation}
  \chi = \frac{N_A N_B\exp[-\beta(\Delta U-\Delta\mu^*_{\mathrm{rxn}})]}{(N_C + 1)(N_D+1)},
\label{eq:forward}
\end{equation}
\begin{equation}
  \chi = \frac{N_C N_D\exp[-\beta(\Delta U+\Delta\mu^*_\mathrm{rxn})]}{(N_A + 1)(N_B + 1)} ,
\label{eq:reverse}
\end{equation}
where $\Delta\mu_{rxn}^*=\Delta\mu_{rxn}+(3/\beta)\ln(\Lambda_C\Lambda_D/(\Lambda_A\Lambda_B))$, $\beta=1/(k_B T)$ is the inverse temperature, $\Lambda_i$ is the de Broglie wavelength of species $i$, 3 is the dimensionality of space, and $\Delta U$ is the change in potential energy from the old to the new configuration.
This trial is implemented in the \mbox{FEASST} class named \texttt{TrialMorph}.

The RxVB move for A + B $\rightleftharpoons$ C + D begins in the same way as the unbiased reaction described above.
In the forward reaction, after A is selected randomly, then if the number of B particles in the reaction volume (RxV) of the chosen A particle, $N^{RxV}_B \neq 0$, one of those B particles in the RxV is randomly chosen.
The RxV is defined as the region between a lower and upper radial distance, $r^{RxV}_l$ and $r^{RxV}_u$, respectively.
Otherwise, the trial is rejected.
The reverse trial randomly chooses a C particle among the $N_C$ particles, and then one of the D particles in the RxV of the chosen C, if $N^{RxV}_D \neq 0$ (otherwise the trial is rejected).
Similarly, local detailed balance yields the following acceptance for the RxVB forward and reverse trials, respectively,
\begin{equation}
  \chi = \frac{N_A N^{RxV}_B \exp[-\beta(\Delta U-\Delta\mu^*_\mathrm{rxn})]}{(N_C + 1)(N^{RxV}_D+1)} ,
\label{eq:forward_acc}
\end{equation}
\begin{equation}
  \chi = \frac{N_C N^{RxV}_D\exp[-\beta(\Delta U+\Delta\mu^*_\mathrm{rxn})]}{(N_A + 1)(N^{RxV}_B + 1)}.
\label{eq:reverse_acc}
\end{equation}
This trial is implemented in the \mbox{FEASST} class named \texttt{TrialRxVB}.

\section{A single and fixed reaction site in a slit pore\label{slitpore_model}}

We first validated the RxVB trial in a slit pore model with a single fixed reaction site that is also publicly available in the first tutorial of the \texttt{rxvb} plugin in the Free Energy and Advanced Sampling Simulation Toolkit (\mbox{FEASST})\cite{hatch_monte_2024} software version 0.25.20.
A slit pore model is a simple geometric construct commonly used in adsorption studies to represent confinement inside porous materials \cite{spohr1998adsorption, ustinov2006adsorption, hirunsit2007effects, shen2014elucidating, mclaren2021examining}.
The model consists of two parallel hard walls separated by a fixed distance (slit width), with molecules confined between the walls and free to move parallel to them.

\begin{figure}
\includegraphics[width=1\columnwidth]{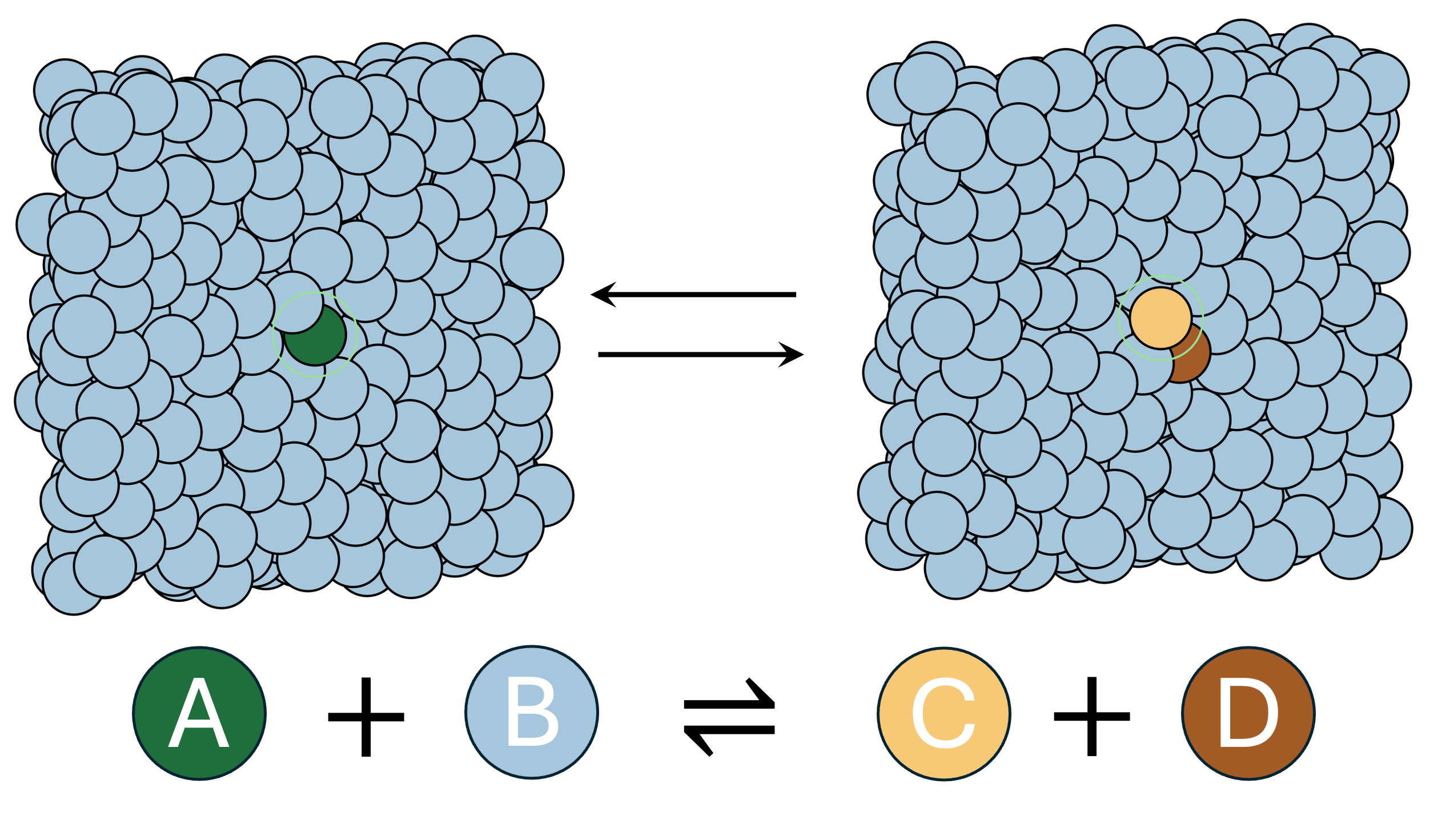}
\caption[Simulation snapshot of slit pore model]
{Schematic of chemisorption in slit pore model. Here, the left side represents the unreacted state with a fixed site A (green) near the wall and fluid B particles (blue), while the right side shows the reacted state, where A-B is replaced by strongly bound products C (yellow) and D (brown). The two panels are independent configurations from the trajectory, so the fluid differs between them; however, the reactive site is fixed at the same position in both.}
\label{slitModelReaction}
\end{figure}

The reaction A$+$B $\rightleftharpoons$ C$+$D was modeled inside the slit pore by fixing a single reactive site (A) just touching the hard wall, while fluid particles (B) were free to move within the pore volume.
Upon reaction, the product species, C and D, replace the A and B pairs, respectively, where C is also fixed to the same position as A.
All particles were represented as hard spheres of unit diameter $\sigma =1$, with interactions described by a square-well potential of range [$\sigma,\,1.05\sigma$].
The reaction volume (RxV) was defined within this attractive shell around the reactive sites A and C.
The interaction strength was set to $\beta \epsilon = 1$ for all pairs, except for the C–D interaction, which was assigned a much stronger attraction of $\beta \epsilon_{CD} = 100$ to favor the formation of products. 
Figure \ref{slitModelReaction} illustrates the schematic representation of the slit pore chemisorption model along with the interaction parameters used. 
The walls acted only as hard confinement, preventing particle overlap beyond the slit boundaries.

All simulations were carried out using the \mbox{FEASST} \cite{hatch_monte_2024} software version 0.25.11 with the \texttt{RxVB} plugin (plugin publicly released in \mbox{FEASST} version 0.25.20), with $10^8$ MC trials for equilibration followed by $10^{10}$ production MC trials at inverse temperature, $\beta=1$.
The slit pore was formed by two parallel hard walls separated by a gap of $4\sigma$, confining 450 fluid particles (B) in one dimension, while periodic boundary conditions of length $12\sigma$ were applied in the other two dimensions.
MC move selection was dominated by particle translations of $\approx 90.9 \%$ of attempted moves, followed by position swap moves $\approx 9.09 \%$, while reaction trials were attempted with a probability of about $0.009 \%$
which is three to four orders of magnitude less frequent than standard moves.
Such an infrequent reaction attempt for the slit pore simulations prevented reaction moves that could be trivially reversed without concern over the orientation of single-site molecules, and therefore too frequent sampling of reactions may not improve configurational sampling.
Position swap moves switched the positions of randomly chosen B and D particles.
The reaction chemical potential difference, $\beta\Delta\mu^*_\mathrm{rxn}=-100$ to sample reactant and product states comparably by compensating for the deep C-D binding well, $\beta\epsilon_{CD}=100$.

\subsection{Evaluating the efficiency of RxVB in a slit pore}\label{slitpore_efficiency}

We consider two complementary efficiency measures to compare the unbiased and biased RxVB MC trials. The first is the efficiency from the block standard deviation, $z_{12}$, as a function of central processing unit (CPU) time, $t$, which captures the combined effect of trial move acceptance and computational overhead \cite{cortes2013influence, schultz2014quantifying, hatch_efficiency_2023, hatch_prefetch_2026}.
Overhead costs include tracking the neighbors of particles, which is not required for the unbiased reaction trial.
Because statistical uncertainty decreases as $\sim 1/t^2$ over long times, the efficiency of simulation 1 relative to simulation 2, $z_{12}$ can be obtained by the ratio of the ordinate intercepts, $b$, with a fit in a log--log plot as\cite{hatch_prefetch_2026}
\begin{equation}
    z_{12}=\exp[2(b_2-b_1)].
\end{equation}
Because $z_{12}$ depends on the CPU time, the results depend upon the details of the hardware and software implementation.

\begin{figure}
\begin{centering}
\includegraphics[width=8cm]{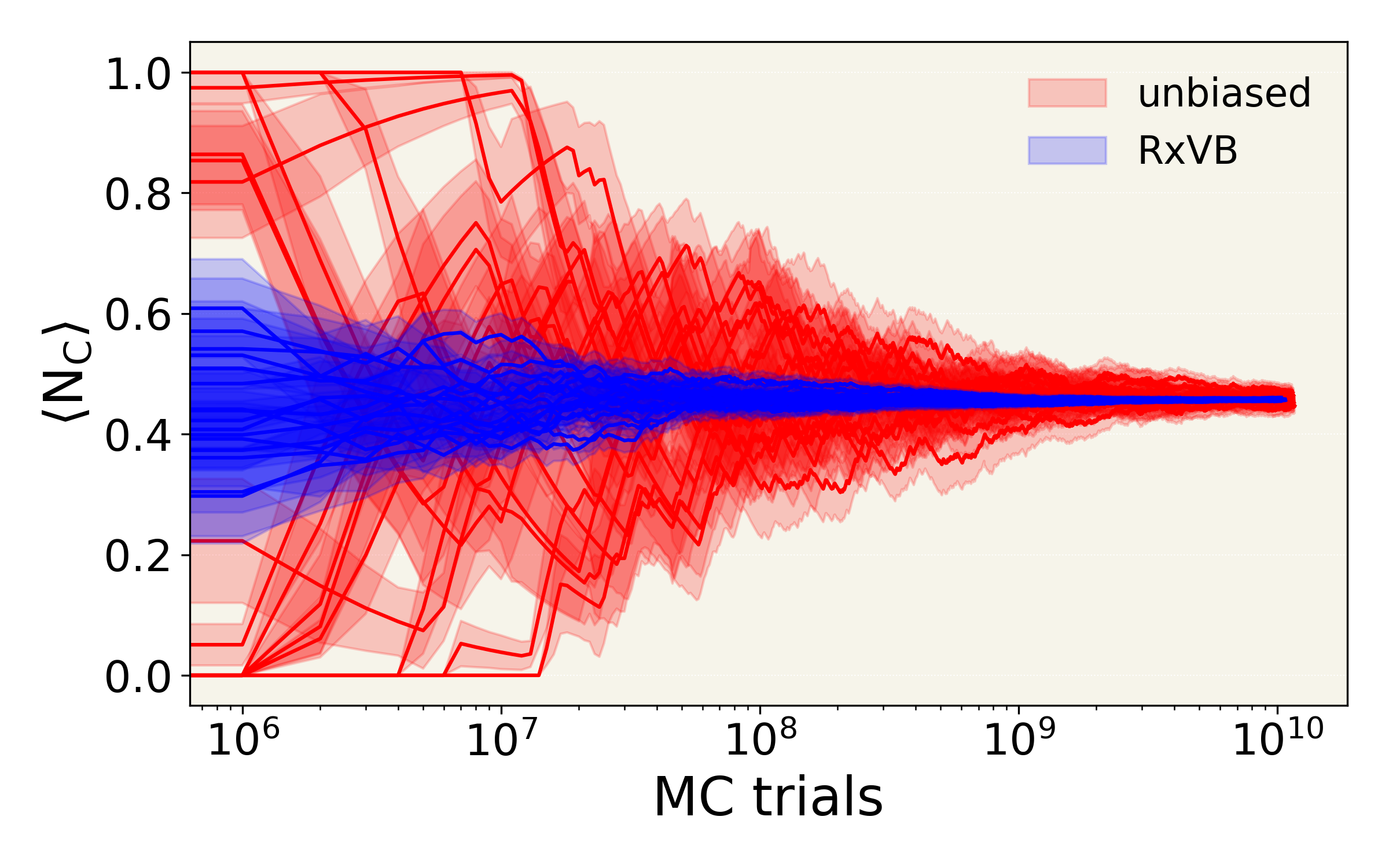}
\caption{Convergence comparison of RxVB and unbiased reactions.
  The ensemble average number of product C particles (inset purple particle) in a slit pore with $N_B=450$ as a function of the number of MC trials with 16 independent simulations using RxVB (blue) and unbiased reactions (red).
  The inset shows a fixed reaction site (inset red D particle) in a fluid of reactant B particles (inset grey) within a slit pore (not shown).
  The shaded transparent regions show the standard deviation of the mean from block averages.\cite{flyvbjerg_error_1989}
}
\label{fig:ncav}
\end{centering}
\end{figure}

\begin{figure}
\begin{centering}
\includegraphics[width=8cm]{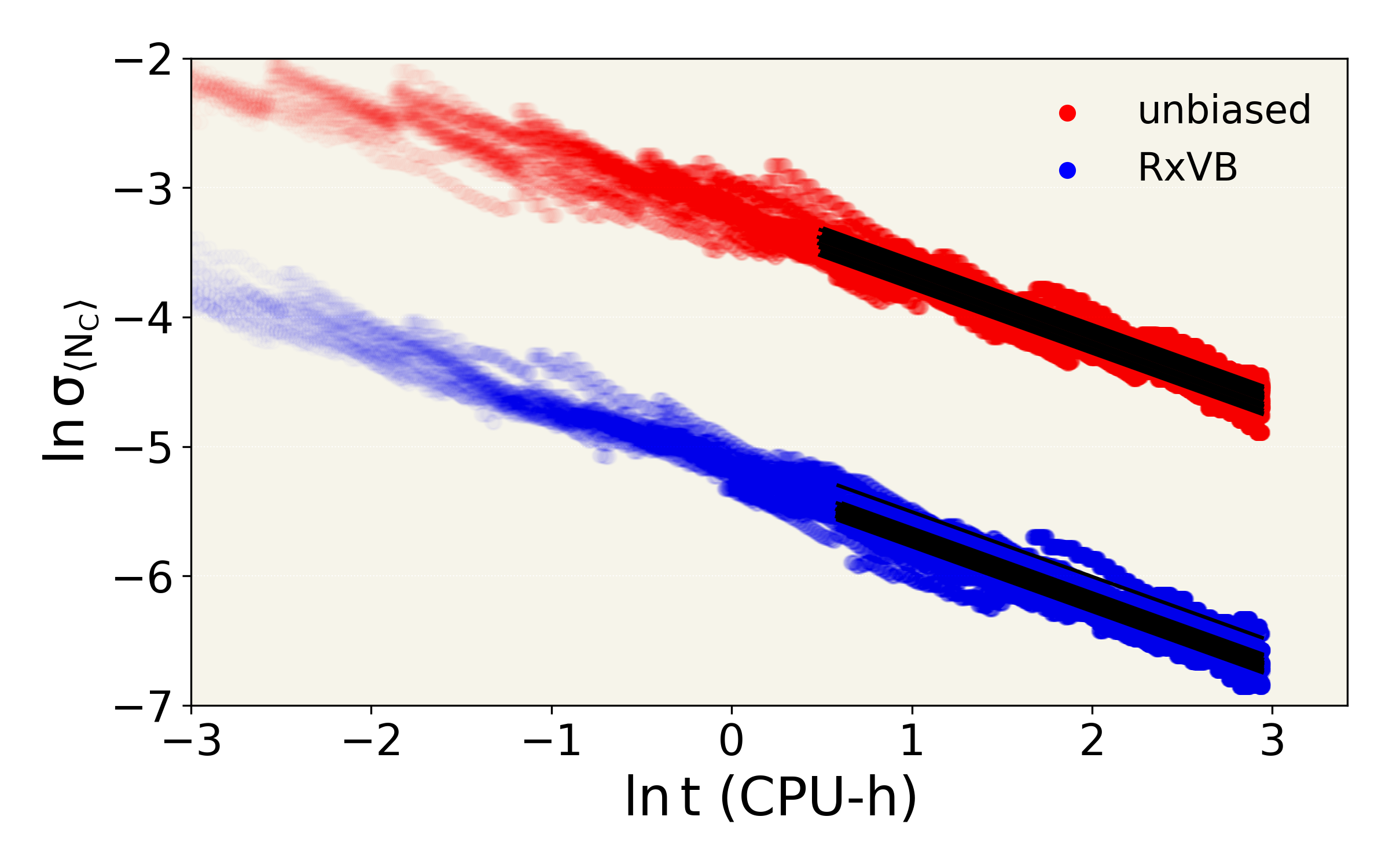}
\caption{The block standard deviation of the mean of the ensemble average number of C particles in the slit pore as a function of CPU time with $N_B=450$ and $16$ independent simulations for RxVB (blue) and unbiased reactions (red).
  Linear fits, shown by the black lines, quantify that RxVB is almost two orders of magnitude more efficient than the unbiased reactions.
}
\label{fig:lnsigncav}
\end{centering}
\end{figure}

The single, fixed reaction site in a slit pore demonstrates the efficiency of the RxVB approach as shown in Fig. \ref{fig:ncav}.
Because of the strong C-D particle interaction, the reaction is highly favored when the reactant A and B particles are in close proximity.
This simple model was developed to demonstrate that RxVB is highly efficient in this case.
The RxVB moves (blue in Fig. \ref{fig:ncav}) converge much more quickly and have smaller standard deviations of the mean than the unbiased reactions (red).
The efficiency, $z$, was quantified by fitting the standard deviation of the mean of the number of C particles, $\sigma_{\langle N_c\rangle}$ with CPU time on a $\ln-\ln$ plot as shown in Fig. \ref{fig:lnsigncav}.
Because a statistically fluctuating quantity is expected to have a linear slope of $-\frac{1}{2}$,\cite{cortes2013influence, schultz2014quantifying} the intercepts were used to defined the efficiency as described by Eq. 4 in Ref. \citenum{hatch_efficiency_2023} are an equilibration time of $5\times 10^6$ trials.
The periodic increase in $\sigma_{\langle N_c\rangle}$ is due to a switchover from $63$ to $32$ blocks.\cite{flyvbjerg_error_1989, hatch_efficiency_2023}

The second efficiency metric considered is the accepted trials ratio, $\text{R}_{\text{acc}}$,
\begin{equation}\label{acc_eq}
R_{\text{acc}}=\frac{\langle N_{\text{acc}}\rangle_{\text{biased}}}{\langle N_{\text{acc}}\rangle_{\text{regular}}},
\end{equation}
where $\langle N_{\text{acc}} \rangle$ is the mean number of accepted reactions per simulation. 
This measure is easier to compute than $z_{12}$ and does not depend on the computer hardware and software optimizations, but $R_\mathrm{acc}$ does not account for the cost of an attempted trial (e.g., due to neighbor-list updates and extra geometric checks in the biased move relative to the unbiased move).

Table~\ref{tab:slit_eff} shows the computed values from the two metrics. The RxVB moves result in approximately 91 times more accepted reactions compared to the unbiased trials, while the $z_{12}=72$ ratio is expected to be lower than $R_\mathrm{acc}$ because of the computational overhead.
Validation of $R_\mathrm{acc}$ against $z_{12}$ allows the use of $R_\mathrm{acc}$ for more computationally expensive simulations (e.g., water) because the calculation of $R_\mathrm{acc}$ requires shorter simulations and less independent simulations than calculation of $z_{12}$.

\begin{table}
\caption{Efficiency metrics for the slit pore. Values at $N_B = 450$ are the mean $\pm$ the standard error of the mean (SEM)  over 16 independent pairs.}
\label{tab:slit_eff}
\begin{ruledtabular}
\begin{tabular}{lcc}
Metric & Value (mean $\pm$ SEM)  \\
\hline
$\text{R}_{\text{acc}}$  & $90.91 \pm 0.91$  \\
$z_{12}$  & $72.4 \pm 3.5$ \\
\end{tabular}
\end{ruledtabular}
\end{table}

\begin{figure}
    \centering
    \begin{subfigure}[b]{0.48\textwidth}
        \centering
        \includegraphics[width=\textwidth]{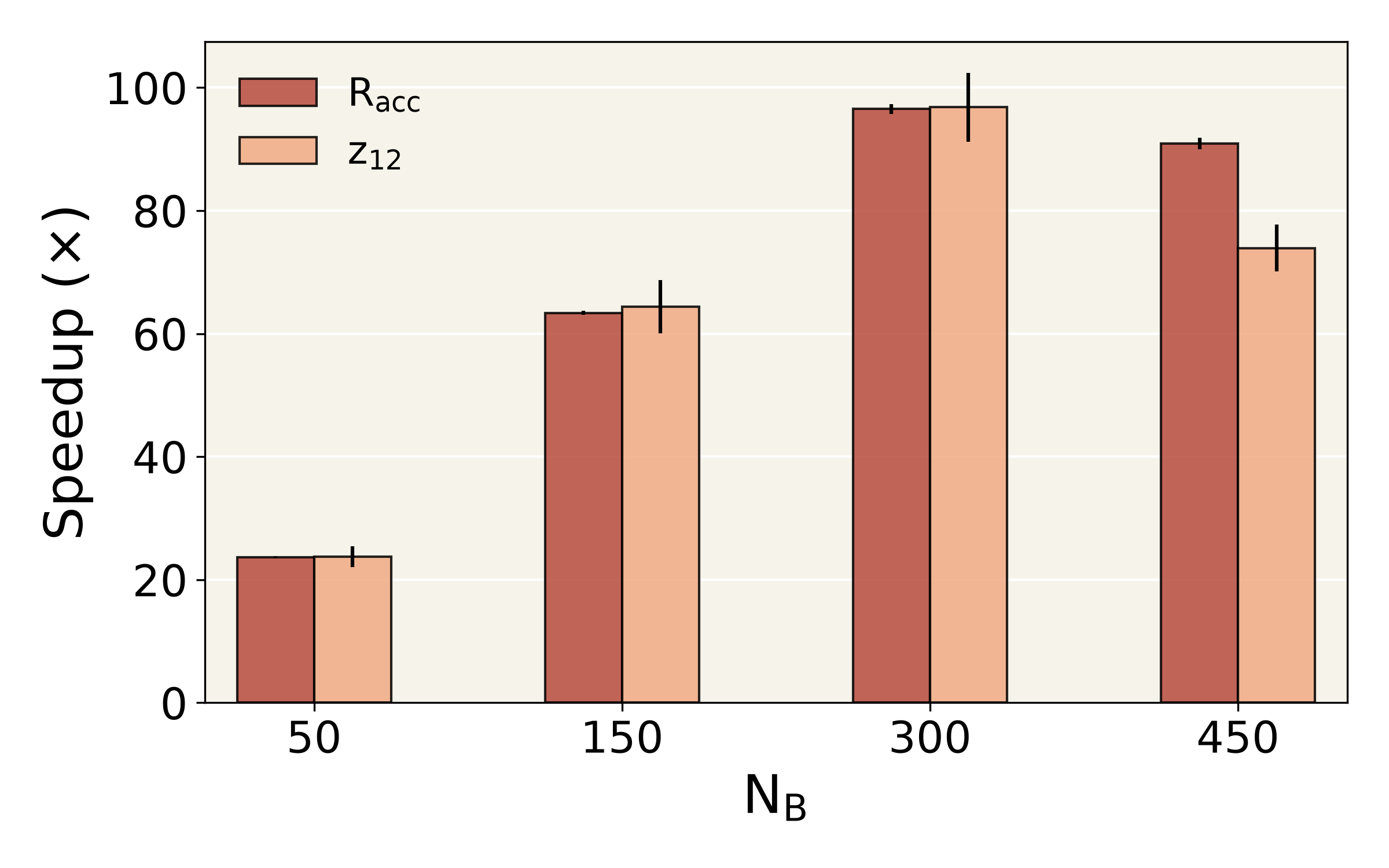} 
        \label{fig:eff_density}
    \end{subfigure}
    \hfill 
    \begin{subfigure}[b]{0.48\textwidth}
        \centering
        \includegraphics[width=\textwidth]{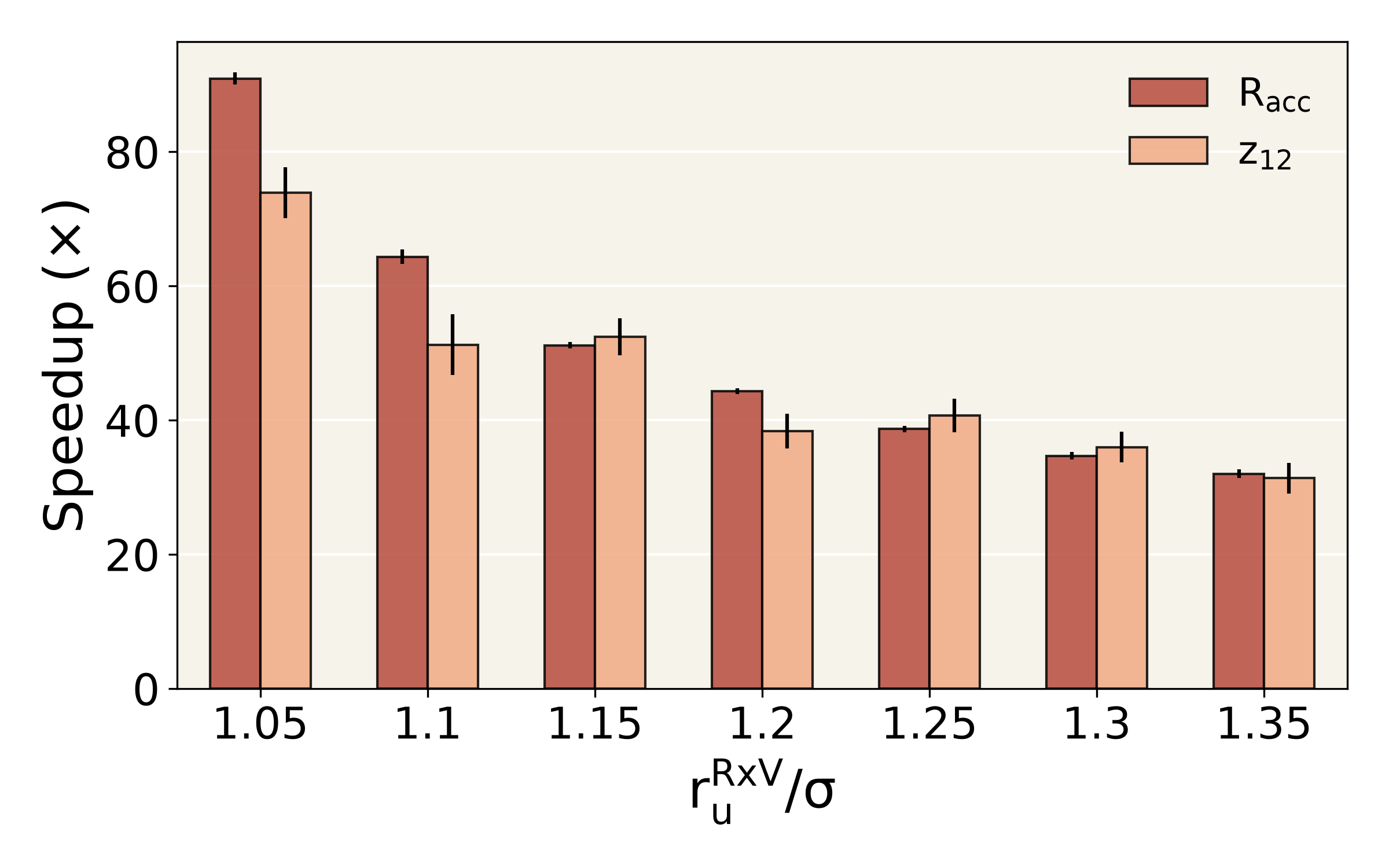} 
        \label{fig:eff_avdist}
    \end{subfigure}
    \caption{Efficiency of the RxVB trial in the slit-pore model ($L = 12\sigma$, slit width $4\sigma$). The speedup ($\times$) on the y-axis is the factor by which RxVB outperforms the unbiased trial. $R_{\mathrm{acc}}$ compares accepted reactions, and $z_{12}$ compares the
    CPU time needed to reach the same statistical uncertainty. Bars show means and error bars denote $\pm$ SEM over 16 independent runs. The top panel illustrates speedup as a function of the number of fluid B particles,
    $N_B$, corresponding to confined densities
    $\rho\sigma^3 = 0.087,\ 0.26,\ 0.52,$ and $0.78125$ , at fixed $r_u^{\mathrm{RxV}} = 1.05\sigma$
    and the bottom panel illustrates speedup as a function of the reaction-volume upper cutoff,
    $r^{\mathrm{RxV}}_u/\sigma$ at fixed $N_B = 450$.}
    \label{fig:slit_efficiency}
\end{figure}

The performance of RxVB moves is expected to depend on how many reacting particles are within the RxV, which is affected by the system size, density, and the RxV geometry.
To quantify these effects, we evaluated multiple densities (top of Figure ~\ref{fig:slit_efficiency}) and RxV distances (bottom of Figure \ref{fig:slit_efficiency}) for the two efficiency metrics introduced above.

As the reaction shell is widened at fixed B density, the efficiency decreases because RxVB includes center-to-center distances between reactant or product pairs that are larger than the square well cut off distance. 
Conversely, higher $N_B$ density boosts efficiency, because biased MC trials benefit from having more candidate partners within the RxV shell, whereas regular trials are less sensitive to this change. 
Since the efficiency metrics follow the same qualitative trends, $R_\mathrm{acc}$ may serve as an indicator of efficiency in cases where $z_{12}$ is much more expensive than $R_{acc}$ to calculation, such as water in confinement.
While the simple geometry of the slit pore model demonstrates that the RxVB trial can be highly effective, other factors might affect these trends when simulating more complex systems.

The rationale for the observed speed up of RxVB compared to unbiased reactions in the slit pore model is as follows.
Due to the large attraction between C and D particles, but the chemical potential which does not favor the creation of C and D, the forward reaction of Eq. \ref{eq:forward} is unlikely to be accepted unless the reacting A and B particles are within the RxV.
While RxVB ensures that is the case, the unbiased reaction must randomly select any B particle and happen to select one in the RxV.
The chances of the unbiased algorithm to select a B particle in the RxV of A is $\langle N_B^{RxV}\rangle/N_B$.
The location of the particles in the reverse reaction (Eq. \ref{eq:reverse}) is not as important because the A and B interactions are not as strong as C and D.
Therefore, the expected slow down of the unbiased reaction is only half of the selection factor $\langle N_B^{RxV}\rangle/N_B$ because it is only important in the forward reaction.
The efficiency qualitatively follows these expectations in the slit pore model, as will be discussed in more detail in comparison to a more complex simulation of water in confinement.

\section{Sampling proton transfer in Br{\o}nsted acid zeolites\label{sec:water}}

\subsection{Construction of a Br{\o}nsted acid zeolite}
\begin{figure}
\includegraphics[width=1\columnwidth]{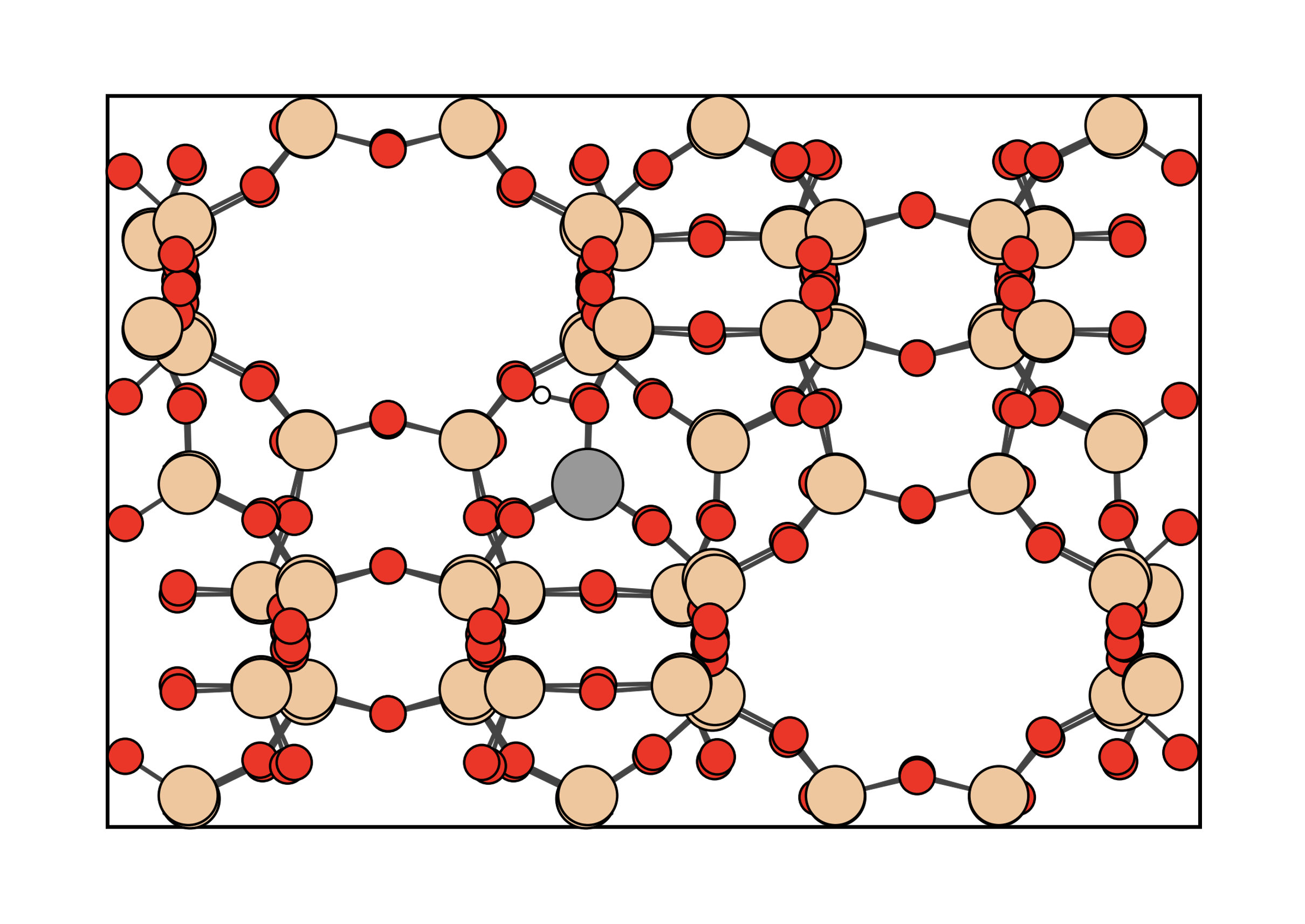}
\caption[Protonated MFI framework with Br{\o}nsted acid site]
{MFI framework with Br{\o}nsted acid site viewed along [100]. A single Al substitution (grey) and its compensating proton (white) on a bridging oxygen (red) within the all-silica MFI framework (Si in tan).}
\label{mfi-protonated}
\end{figure}

Next, we applied the RxVB move to sampling acid-base reactions of water within Br{\o}nsted acid zeolites. 
An MFI framework consisting of a single unit cell was constructed with a fixed protonated site and a Si/Al ratio of 95, as illustrated in Figure \ref{mfi-protonated}.
The starting structure was an all-silica MFI, wrapped to unit cell dimensions, with TraPPE-Zeo Lennard-Jones parameters assigned to framework atoms \cite{bai2013trappe}. 
A single Br{\o}nsted acid site was introduced by substituting one Si (tan) with Al (grey) at a T--site, yielding a net framework charge of $-1$. 
The compensating proton ($+1$, white) was placed on a neighboring bridging oxygen (red), and its position was generated geometrically.
The ring of framework oxygens around the chosen site was used to estimate the pore center, and the proton was then positioned at a fixed O–H bond length of $1\,\text{\AA}$ along the line toward that center. 

\subsection{Monte Carlo simulation details}
\par
\begin{figure*}
\begin{centering}
\includegraphics[width=1\textwidth]{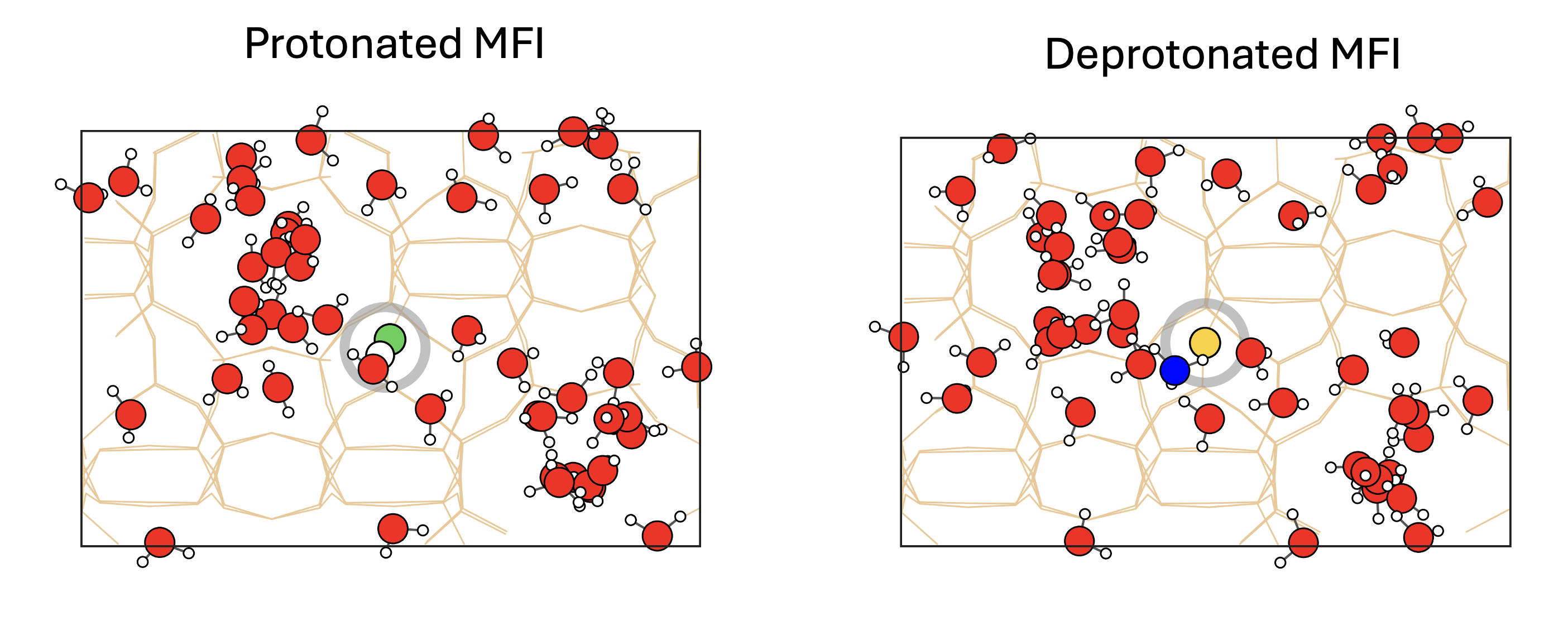}
\caption{Water adsorption in MFI framework with Br{\o}nsted acid site. 
Simulation snapshots before (left) and after (right) chemisorption. Framework bonds (Si–O and Al–O) are drawn in orange for context. Atom sizes are scaled according to their LJ $\upsigma$ values. Shaded grey annuli indicate the RxV inner and outer sampling radii, centered on the target active-site oxygen atoms (green in the unreacted state, yellow in the reacted state). The oxygen atom of hydronium is shown in blue, while that of water molecules is shown in red.}
\label{MFI_chemisorption}
\end{centering}
\end{figure*}

 The simulations used five particle types: (i) a rigid, nonreactive MFI framework, 
(ii) the protonated acid site (Al--O--H), (iii) the deprotonated site (Al--O$^{-}$), 
(iv) SPC/E water \cite{berendsen1987missing}, and (v) the H$_3$O$^{+}$ of Bonthuis et al. \cite{bonthuis_optimization_2016} 
Types (ii)--(iii) and (iv)--(v) form coupled identity pairs for the reaction,
\begin{equation}
\text{framework-}\mathrm{OH} + \mathrm{H_2O}
\;\rightleftharpoons\;
\text{framework-}\mathrm{O^-} + \mathrm{H_3O^+}
\end{equation}
Figure \ref{MFI_chemisorption} illustrates simulation snapshots in the reacted and unreacted states in the MFI framework for the 50-water system.

\par All force field parameters are listed in Table \ref{FFparamsChap3}. Lennard-Jones (LJ) potentials were used for short-range van der Waals interactions and Coulomb potentials for electrostatics, both with a spherical cutoff of 6.5\,\AA.
The cutoff was set to about half the shortest unit cell dimension ($c = 13.383$\,\AA) so as to satisfy the minimum image convention and enable long benchmark simulations with minimal system sizes.
LJ interactions were force-shifted to zero at the cutoff, and long-range electrostatics were treated by Ewald summation.
Trial selection probabilities mirrored the slit pore model setup, with rigid-body translations and rotations of mobile particles dominating at $\approx 95 \%$, position swap moves at $\approx 4.5 \%$, and reaction trials at $\approx 0.5 \%$.
Reaction trials were attempted more frequently than in the slit pore model, but we did not determine an optimal reaction attempt probability, which may depend a number of algorithmic and thermodynamic parameters.
Position swap moves exchanged the oxygen positions of a randomly selected water molecule with the hydronium, and then randomly oriented both molecules about their oxygen sites.
For reaction trials, all hydrogens involved in the reaction (e.g., H$_2$O, H$_3$O and the framework-OH) were randomly oriented about their oxygen, as described in Sec.~\ref{sec:acceptance}.
Although utilizing the old positions of the hydrogens to prevent disruption of the hydrogen bond network may improve sampling, such a method is difficult to implement for models of water with rigid hydrogen bond lengths and angles.

\par Each state point was sampled with 32 independent replicas: 16 with biased moves and 16 with unbiased moves across eight water loadings (1w, 2w, 3w, 4w, 6w, 8w, 10w, 50w, where w denotes the number of water molecules) and chemical potentials between ($-700$ and $-300$) kJ$/$mol.
Simulations were frequency checkpointed and restarted for very long simulations on high performance computing clusters, reaching approximately $1.5\text{--}2 \times 10^9$ trials per replica over roughly five months.
The production stage was defined as the final third of the simulation trajectories.

\subsection{Optimizing the reaction volume}

\par
\begin{figure}[H]
\begin{centering}
\includegraphics[width=0.475\textwidth]{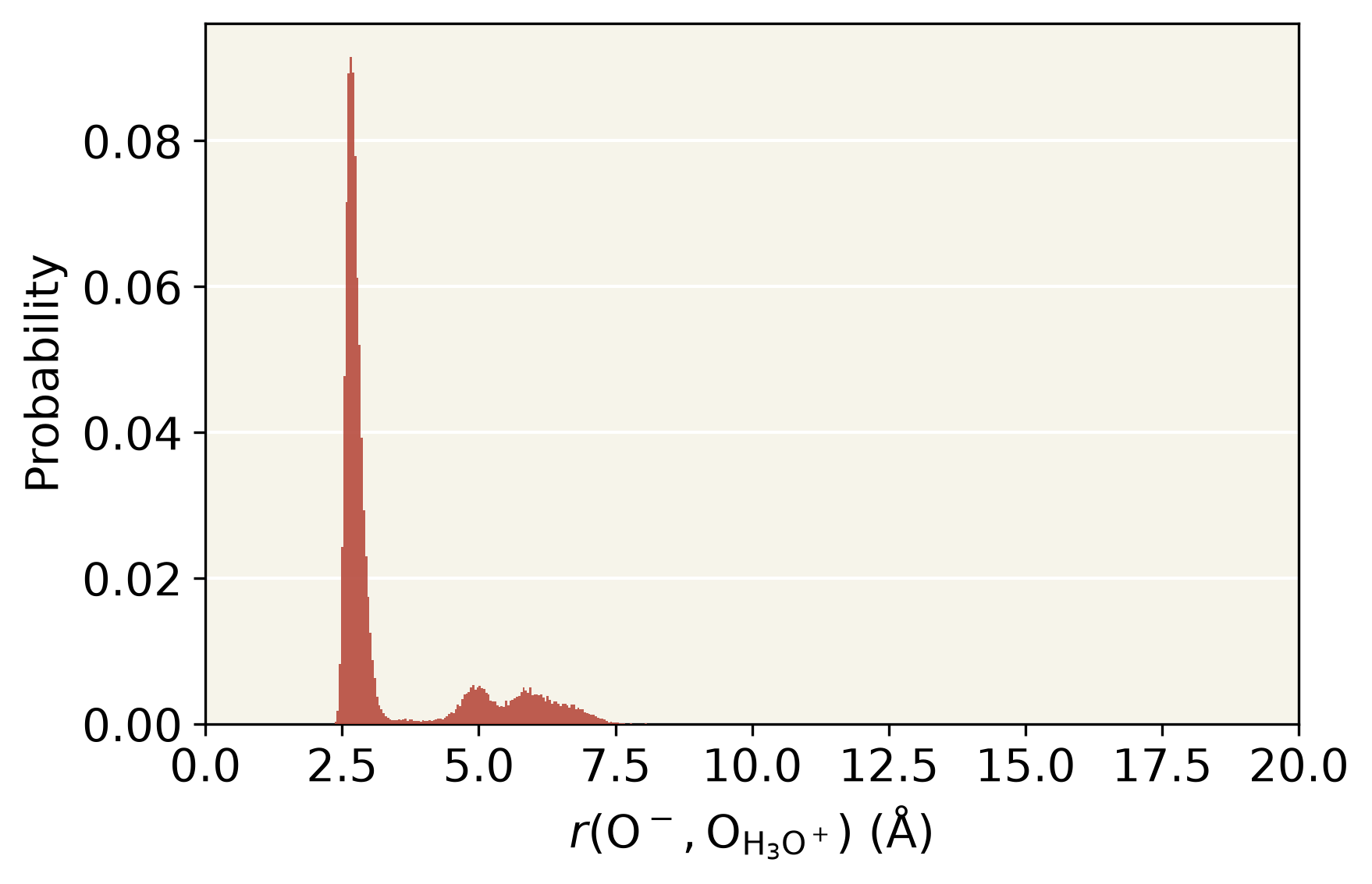}
\caption{Distribution of hydronium oxygen distance (${\mathrm{H_3O^+}}$) from the acid site oxygen in MFI, computed with the minimum image convention over all equilibrium frames in which both product species were present (the reacted state) with 50 water molecules and $\upmu = -550$\,kJ/mol.}
\label{mfi_product_dist}
\end{centering}
\end{figure}

 Before using RxVB, we should determine an optimal definition for the reaction volume.
We base the reaction volume on the distances between the oxygens of the reacting species.
The RxVB move is intended to improve sampling by preferentially proposing reactions where a candidate water occupies the reaction-volume (RxV) shell around the reactive site.
The inner and outer RxV distances were chosen so that the RxV shell encompasses the first solvation shell of the reactive site oxygen.
As the O--O Lennard-Jones interactions of the SPC/E water model overlap significantly ($\sigma \approx 3.17$\,\AA) \cite{berendsen1987missing}, we adopted a relatively small inner distance of $r^{RxV}_l=2.4$\,\AA\ together with an outer distance of $r^{RxV}_u=3.7$\,\AA.
If the RxV shell is too small, there may not be enough water molecules available for the reaction; conversely, if it is too large, RxVB may have a diminished effect.
We further examined the distance distribution of the selected product pair (deprotonated site oxygen and the hydronium oxygen (O$^-$--O$_{\mathrm{H_3O^+}}$)) to confirm that this shell is appropriate for the MFI--water system.
As shown in Figure \ref{mfi_product_dist}, the O$^-$--O$_{\mathrm{H_3O^+}}$ separation is sharply peaked near 2.7\,\AA, corresponding to a contact ion pair in which the hydronium sits directly adjacent to the deprotonated acid site, with a smaller secondary population between roughly 5 and 7\,\AA\ representing configurations one solvation shell away.
The chosen RxV shell of 2.4--3.7\,\AA\ encloses this dominant contact-pair population, confirming that it captures the configurations most relevant to the reaction.

\subsection{Determining the chemical potential for equal probability of reactant or product}\label{mu_sweep}

The efficiency of a reacting trial is difficult to compute if the simulation conditions are such that most configurations include only the reacted products, or most configurations include only the reactants.
The probability that a site is reacted is influenced by the chemical potentials of all of the species.
In this section, we attempt to find the chemical potential at which roughly half of the configurations contain the products.

\par
\begin{figure}[htbp]
\centering
    \begin{subfigure}[b]{\linewidth}
        \centering
        \includegraphics[width=\linewidth]{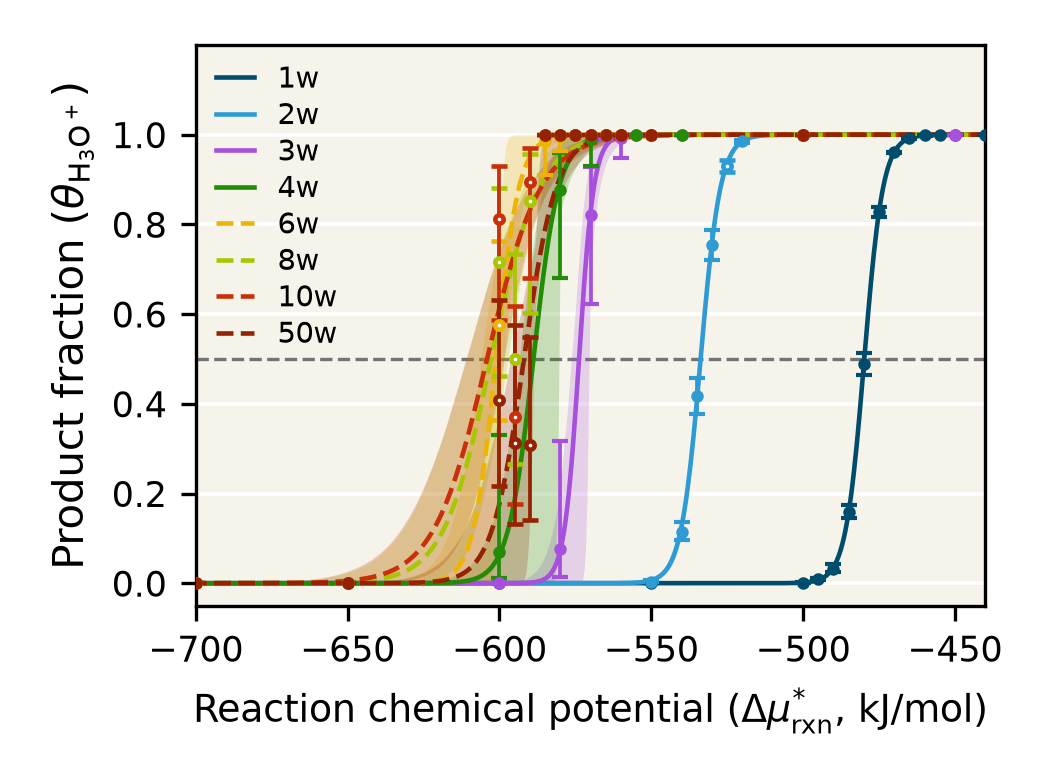}
        \label{fig:biased}
    \end{subfigure}
    \vspace{3pt}
    \begin{subfigure}[b]{\linewidth}
        \centering
        \includegraphics[width=\linewidth]{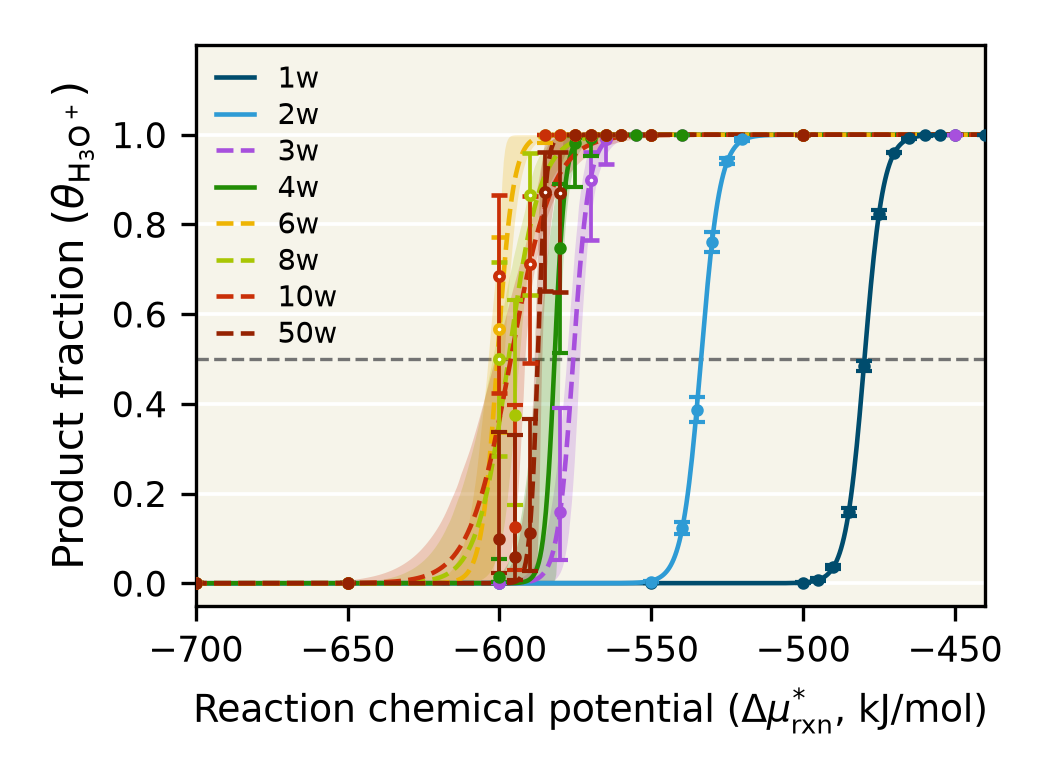}
        \label{fig:unbiased}
    \end{subfigure}
    \caption{Product fraction of hydronium, $\theta_{\mathrm{H_3O^+}}$, as a function of the reaction chemical-potential difference $\Delta\mu^{*}_{\mathrm{rxn}}$ for the protonated single-unit MFI framework at water loadings of 1w--50w water molecules per unit cell, using the RxVB (top) and the unbiased (bottom) reaction trial. Points are means over the 16 replicas of each scheme, taken from the final third of the trajectories. Error bars are 95 \% confidence intervals on each point, computed on the logit scale so they respect the physical bounds $[0,1]$. The band is a bootstrap: the curve is refitted 1000 times to randomly redrawn sets of the same 16 replicas, and the band spans the middle 95 \% of those fits. Filled symbols and solid curves indicate that the final third of the trajectory agrees with the middle third, within two standard errors across replicas or to better than 0.02 in $\theta$; open symbols and dotted curves indicate a larger difference. The grey dashed line marks $\theta = 0.5$, which defines the midpoint $\upmu^{*}$.}
    \label{fig:mu_sigmoids}
\end{figure}

 The H$_3$O$^{+}$ chemical potential dependence was investigated by performing a sweep of simulations with $\upmu \in [-700,-300]$ kJ/mol, to find the $\upmu$ where reactants and products have approximately the same probability.
We first initialized the system in the grand canonical ensemble, and then for each $\upmu$, simulations were conducted at fixed $T = 373$ K and $V$ with semi-grand identity changes (H$_2$O $\leftrightarrow$ H$_3$O$^+$, coupled to OH $\leftrightarrow$ O$^-$).
The resulting product fraction ($\theta_{\mathrm{H_3O^+}}$) curves are sigmoidal (Fig.~\ref{fig:mu_sigmoids}), and the midpoint $\upmu^*$ is the chemical potential at which the reactant and product states are equally populated.
The plotted points are means over the 16 replicas for each scheme (biased and unbiased), with error bars and shaded bands showing 95 \% confidence intervals.

\par For the systems containing one and two water molecules, the two schemes agree closely: $\upmu^*$ is $\approx -480$ kJ/mol at one water and $\approx -533$ kJ/mol at two.
This agreement is a check that RxVB trial satisfies detailed balance in a zeolite framework, complementing the slit-pore validations of Section \ref{slitpore_model}.
At three water molecules and above, $\upmu^*$ cannot be reliably determined; these state points average one to three accepted reactions per replica, so the plotted $\theta_{\mathrm{H_3O^+}}$ reflects how far the relaxation progressed rather than an equilibrium average.

\par Two features of the data indicate the uncertainty in the reported chemical potential estimates.
First, the drift between the middle and final thirds of the trajectories is positive at every under-sampled state point at which any reaction was accepted, so the $\theta_{\mathrm{H_3O^+}}$ is under-estimated.
As the $\theta_{\mathrm{H_3O^+}}$ increases with less negative $\upmu$, the fitted $\upmu^*$ is correspondingly an upper bound, and the true value lies at more negative $\upmu$.
Second, the discrepancy between the two schemes grows with loading, from 0.1--0.2 kJ/mol at one and two waters to 6--9 kJ/mol at eight and ten, providing a data-driven estimate of the systematic error, since at equilibrium the two schemes must agree.

\par The midpoint shifts to more negative $\upmu$ as water is added, from $\approx -480$ kJ/mol at one water to $\approx -604$ kJ/mol at ten.
Since $\upmu^*$ measures the penalty required to hold the reaction at equal populations, this indicates that a higher local water concentration intrinsically favors deprotonation of the acid site in confinement.

\par The water-loading dependence observed here is qualitatively consistent with prior first-principles studies of Br{\o}nsted-site zeolites in water \cite{markovitch2007structure, grifoni2021confinement, baiFirstPrinciplesGrandCanonicalSimulations2021, hack2023proton, liu2024water}.
Grifoni et al. \cite{grifoni2021confinement} used ab initio molecular dynamics with enhanced sampling and showed a progressive shift from a framework-bound proton to a fully solvated hydronium ion as the water content near the acid site increased, with complete solvation requiring more than roughly three water molecules.
Bai et al. \cite{baiFirstPrinciplesGrandCanonicalSimulations2021} reached a similar conclusion for the same H-ZSM-5 (MFI) framework using first-principles grand-canonical Monte Carlo, finding that the proton dissociates readily into the adsorbed water network and that the extent of water clustering is limited by framework confinement.
However, both approaches evaluate the proton-transfer degrees of freedom on a first-principles potential-energy surface and sample the deprotonation through dynamics.
Here we instead sample the deprotonation equilibrium explicitly and classically, using a RxVB move.
As we impose $\Delta\upmu_{\text{rxn}}$, these simulations do not test the reported three-water threshold directly and only show the trend that the chemical potential required to reach equal populations falls monotonically as water is added.
The identity-switch move also samples only the protonated and deprotonated end states and does not resolve the shared-proton (Z\"undel-like) intermediate captured by the first-principles free-energy surfaces.

\subsection{Efficiency in the Br{\o}nsted acid MFI-water system}

\par The efficiency of RxVB was quantified with the accepted trials ratio $R_\mathrm{acc}$ defined in Section~\ref{slitpore_efficiency}. The block standard deviation efficiency $z_{12}$ was not computed because it required too much CPU time to converge, even though some simulations were run for over five months.
At the under-sampled state points, the acid site changes protonation state only a few times at most (if at all) over the entire run, so $N_\mathrm{acc}$
(Table~\ref{av_occupancy}) is small.
The small counts at higher loadings in MFI are consistent with literature studies that report slow convergence of water in nanoporous frameworks \cite{zhang2017computational, formalik2025small}.

\begin{table*}
\begin{ruledtabular}
\begin{tabular}{lccrrcc}
system & $N_B$ & $\langle N^{RxV}_B \mid N^{RxV}_B \geq 1 \rangle$ & $N_\mathrm{acc}$ biased & $N_\mathrm{acc}$ unbiased & $R_\mathrm{acc}$ & $z_\mathrm{max}$ \\
\hline
slit pore & 50 & 1.045 &  45\,520 & 1\,925 & $23.64 \pm 0.17$ & 23.9 \\
slit pore & 150 & 1.162 &  94\,439 & 1\,490 & $63.40 \pm 0.31$ & 64.5 \\
slit pore & 300 & 1.479 & 106\,168 & 1\,100 & $96.54 \pm 0.78$ & 101.4 \\
slit pore & 450 & 2.529 &  61\,475 &    676 & $90.91 \pm 0.91$ & 89.0 \\
MFI & 1 & 1.000 &      417 &    831 & $0.50 \pm 0.01$ & 0.50 \\
MFI & 2 & 1.053 &      127 &    210 & $0.60 \pm 0.01$ & 0.95 \\
MFI & 3 & 1.178 &      3.3 &    3.3 & $1.00 \pm 0.27$ & 1.27 \\
MFI & 4 & 1.299 &      1.8 &    2.1 & $0.88 \pm 0.19$ & 1.54 \\
MFI & 6 & 1.584 &      1.2 &    0.9 & $1.27 \pm 0.45$ & 1.89 \\
MFI & 8 & 1.613 &      0.9 &    0.9 & $0.93 \pm 0.11$ & 2.48 \\
MFI & 10 & 1.605 &      1.1 &    0.7 & $1.64 \pm 0.41$ & 3.12 \\
MFI & 50 & 2.501 &      1.0 &    0.9 & $1.07 \pm 0.07$ & 10.0 \\
\end{tabular}
\caption{Reaction-volume occupancy and speedup for slit-pore and MFI systems. $N_B$ is the number of reactant molecules present; for MFI this is
the water loading. $\langle N^{RxV}_B \mid N^{RxV}_B \geq 1 \rangle$ is the mean
number of reactant molecules in the RxV shell, measured in the reactant state and
conditioned on the shell being occupied; standard errors are below 0.04
throughout and are not shown. $N_\mathrm{acc}$ is the number of accepted reaction
moves per replica, averaged over the 16 replicas of each scheme; each accepted
move is a transition between protonation states. $R_\mathrm{acc}$ is the observed
speedup, the ratio of the two $N_\mathrm{acc}$ columns, with the uncertainty
propagated from the standard error of each mean and $z_\mathrm{upper}$ follows
from Eq.~\ref{z_conditional}.}
\label{av_occupancy}
\end{ruledtabular}
\end{table*}

\par The $R_\mathrm{acc}$ ratio ranges from $0.50 \pm 0.01$ at one water to $1.64 \pm 0.41$ at ten and $1.07 \pm 0.07$ at fifty, remaining of order unity throughout.
This behavior follows from how many reactant molecules occupy the reaction volume.
The shell is empty most of the time at low density, and an average that includes those configurations predicts the speedup should fall as density rises, which runs against the trend in Figure \ref{fig:slit_efficiency}.
When the reaction volume is occupied (the only case in which the biased trial can propose at all), the unbiased identity switch selects a partner uniformly from the $N_B$ reactant molecules present and finds one inside the shell with probability $\langle N^{RxV}_B \mid N^{RxV}_B \geq 1 \rangle / N_B$, whereas the biased trial selects among the molecules already in the shell and therefore always does.
Comparing the two gives
\begin{equation}
z_\mathrm{max} \;=\; \frac{N_B}{2\,\langle N^{RxV}_B \mid N^{RxV}_B \geq 1 \rangle},
\label{z_conditional}
\end{equation}
where the factor of two arises because the bias only helps in the forward direction.

\begin{figure}
\begin{centering}
\includegraphics[width=1\columnwidth]{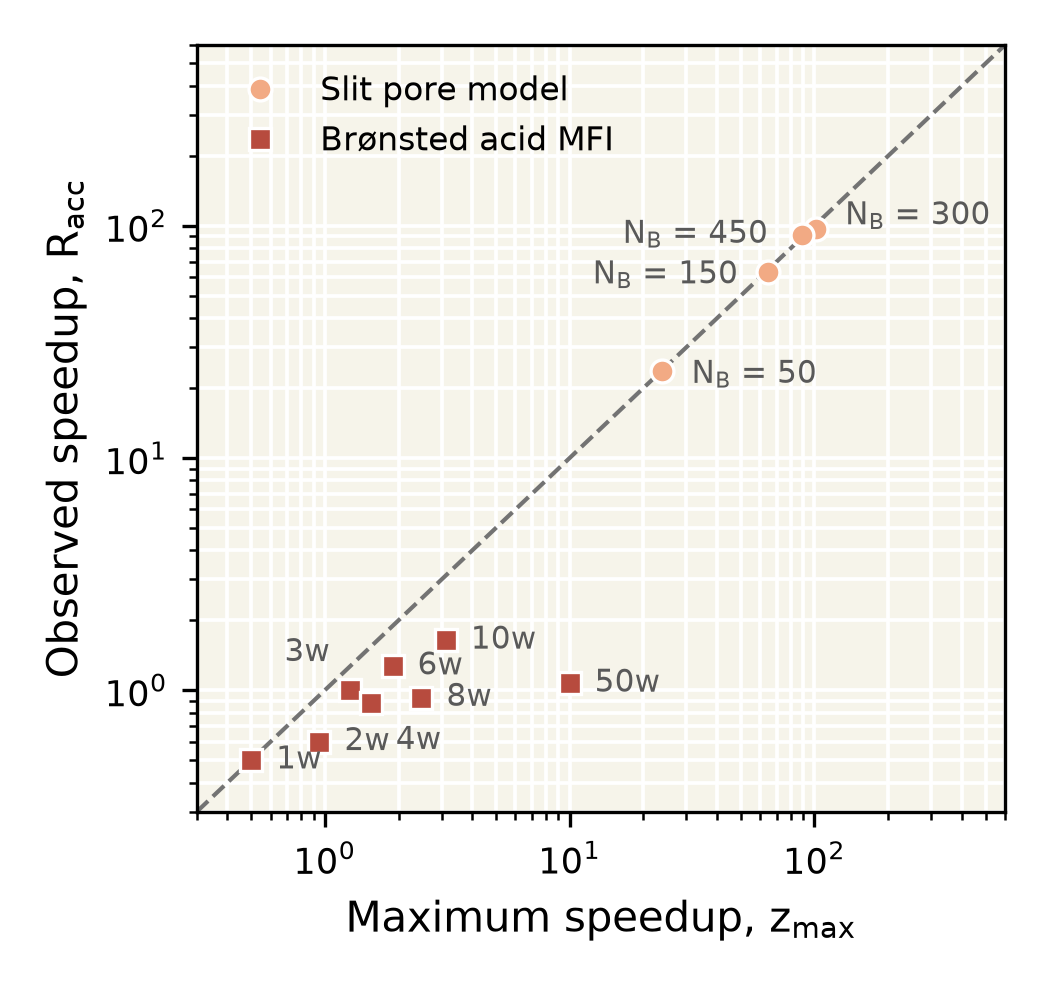}
\caption{The accepted-trials ratio $R_\mathrm{acc}$ compared to the maximum speedup, $z_\mathrm{max}$ predicted by Eq.~\ref{z_conditional}, for the slit-pore model at four reactant densities and for MFI at eight water loadings ($n$w denotes $n$ water molecules per unit cell, from 1w to 50w). The dashed line is parity.}
\label{av_parity}
\end{centering}
\end{figure}

Figure~\ref{av_parity} shows the observed against the maximum speedup for both systems.
For the slit pore, where observed speedups span 24 to 97, Eq. \ref{z_conditional} reproduces $R_\mathrm{acc}$ to within about 5 \%.
The slit pore results validate the expression for the maximum speedup, $z_\mathrm{max}$ of the RxVB move when molecular orientations do not affect the energetics or trial acceptances.
In MFI, the one-water (1w) system serves to validate the RxVB move and $R_\mathrm{acc}$ metric.
Because RxVB cannot possibly be more efficient than the unbiased reaction if there is only one water to select, the 1w simulation is the most thoroughly sampled state point studied, and agrees with the prediction to better than 1 \% ($z_\mathrm{max} = 0.50$, $R_\mathrm{acc} = 0.50 \pm 0.01$), as expected.
The acceptance ratio of the two water (2w) system, $R_\mathrm{acc} = 0.60 \pm 0.01$ falls below the maximum speedup, $z_\mathrm{max} = 0.95$.
At higher loadings, the measured ratios fall increasingly far below the prediction.
With three or more water molecules, each replica accepts on average three reactions or fewer (Table \ref{av_occupancy}).
The measured values are lower bounds.
The RxVB move did not appear to lower the free-energy barrier to reorganizing the hydrogen-bond network.
Sampling that barrier may require additional configuration-bias methods for hydrogen placement, as opposed to randomly orienting the reactants and products.

\par The conditional occupancy $\langle N^{RxV}_B \mid N^{RxV}_B \geq 1 \rangle$ is 1.0--2.5 in both the MFI and the slit pore, indicating that the RxV shell holds essentially the same number of molecules in both systems (Table \ref{av_occupancy}).
The ratio $N_B / \langle N^{RxV}B \mid N^{RxV}B \geq 1 \rangle$, which is twice $z_\mathrm{max}$, ranges from 48 to 203 across the slit-pore densities but only from 1 to 20 across the MFI loadings (Table~\ref{av_occupancy}).
At physically accessible water loadings in an MFI unit cell, geometric pre-selection offers less benefit compared to the slit pore when there is less water outside the RxV.
This gives a criterion for when RxVB moves shows how efficiency scales with the number of reactant molecules outside the shell, so RxVB may increase efficiency when
$N_B \gg \langle N^{RxV}_B \mid N^{RxV}_B \geq 1 \rangle$ and offers negligible advantage when the two are comparable.
Larger framework sizes or higher reactant loadings would benefit more from RxVB than the single unit cell studied here.
\section{Conclusion\label{sec:conclusion}}

We combined an identity-switch reaction move with volume bias to create a reaction volume bias (RxVB) Monte Carlo trial for molecular simulations, implemented it in the open-source \mbox{FEASST} package,\cite{hatch_monte_2024} and validated RxVB against unbiased reactions in a slit pore model with a square-well fluid and in a Br{\o}nsted acid MFI framework with a rigid point-charge model of water.
We demonstrate how the RxVB trial may improve sampling by proposing reactions only between reactants that lie within a prescribed volume around the reactive site.
An expression for the maximum RxVB speedup, Eq.~\ref{z_conditional}, is derived from the relative amount of reactants outside the reaction volume relative to inside.

In a single-site slit-pore model, the RxVB trial may accept up to 91 times more reactions than the unbiased reaction at the highest density.
In a Br{\o}nsted acid MFI framework with a single unit cell and up to 50 water molecules, the RxVB did not result in a measurable speedup compared to the unbiased reaction.
A RxVB trial may increase efficiency by excluding the reactant molecules that lie outside the reaction volume, but in a single MFI unit cell at the loadings considered in this work, there are few to exclude.
Although our water-MFI simulations did not show increased speedup, these results validate the RxVB move and the maximum speedup based on occupancy.
Additionally, improving sampling at higher loadings may require additional configuration-bias methods to cross the hydrogen-bond reorganization barriers, because the current RxVB implementation randomly oriented each of the hydrogens involved in the reaction subject to their rigid bond length constraints with oxygen.

Within these limits, the classical reaction ensemble is inexpensive compared with first-principles approaches. Ab initio molecular dynamics and first-principles grand-canonical Monte Carlo are restricted to small cells and
short trajectories by the cost of evaluating the electronic potential-energy surface, \cite{fetisovUnderstandingReactiveAdsorption2018, fetisovFirstPrinciplesMonte2018, baiFirstPrinciplesGrandCanonicalSimulations2021} whereas the method used here samples chemisorption up to the full pore capacity of the single-unit MFI framework, on the order of 40--50 water molecules,\cite{baiFirstPrinciplesGrandCanonicalSimulations2021} at a small fraction of that cost.
This makes it practical to map the water-loading dependence of Br{\o}nsted-site protonation across a range of loadings that would be expensive to reach from a first-principles approach.
\section{Supporting Information}

See Table~\ref{FFparamsChap3} for a summary of the model parameters.

\begin{table*}
\centering
\small
\setlength{\tabcolsep}{4pt}
\begin{tabular}{clccccc}
\rowcolor[HTML]{EFEFEF} 
\multicolumn{7}{c}{\cellcolor[HTML]{EFEFEF}parameters for non bonded potentials} \\ \specialrule{.1em}{.1em}{.1em} 
\rowcolor[HTML]{EFEFEF} 
\multicolumn{2}{c}{\cellcolor[HTML]{EFEFEF}type} & pseudo atom & $\sigma$ [Å] & $\epsilon$ [kJ/mol] & $q$ [e] & refs \\
\multicolumn{2}{c}{hydronium} & O & 3.1 & 0.8 & $-$1.4 & \cite{bonthuis_optimization_2016} \\
\multicolumn{2}{c}{hydronium} & H & 0 & 0 & 0.8 & \cite{bonthuis_optimization_2016} \\
\multicolumn{2}{c}{water} & O & 3.17 & 0.6502 & $-$0.8476 & \cite{berendsen1987missing} \\
\multicolumn{2}{c}{water} & H & 0 & 0 & 0.4238 & \cite{berendsen1987missing} \\
\multicolumn{2}{c}{zeolite} & Si & 2.30 & 0.1829 & 1.50 & \cite{bai2013trappe} \\
\multicolumn{2}{c}{zeolite} & O & 3.30 & 0.4407 & $-$0.75 & \cite{bai2013trappe} \\
\multicolumn{2}{c}{zeolite} & Al & 2.30 & 0.1829 & 0.5 & \cite{bai2013trappe}, this work$^{a}$ \\
\multicolumn{2}{c}{zeolite} & H & 0 & 0 & 1 & this work$^{a}$ \\
\rowcolor[HTML]{EFEFEF} 
\multicolumn{7}{c}{\cellcolor[HTML]{EFEFEF}parameters for bonded potentials} \\ \specialrule{.1em}{.1em}{.1em} 
\rowcolor[HTML]{EFEFEF} 
\multicolumn{2}{c}{\cellcolor[HTML]{EFEFEF}fixed bond} & \multicolumn{4}{c}{\cellcolor[HTML]{EFEFEF}length {[}Å{]}} & refs \\
\multicolumn{2}{c}{O--H (hydronium)} & \multicolumn{4}{c}{0.98} & \cite{bonthuis_optimization_2016} \\
\multicolumn{2}{c}{O--H (water)} & \multicolumn{4}{c}{1.0} & \cite{berendsen1987missing} \\
\rowcolor[HTML]{EFEFEF} 
\multicolumn{2}{c}{\cellcolor[HTML]{EFEFEF}bend angle} & \multicolumn{4}{c}{\cellcolor[HTML]{EFEFEF}$\theta$ {[}deg{]}} & refs \\
\multicolumn{2}{c}{H--O--H (hydronium)} & \multicolumn{4}{c}{111.4} & \cite{bonthuis_optimization_2016} \\
\multicolumn{2}{c}{H--O--H (water)} & \multicolumn{4}{c}{109.47} & \cite{berendsen1987missing} \\
\rowcolor[HTML]{EFEFEF} 
\multicolumn{7}{c}{\cellcolor[HTML]{EFEFEF}parameters for the reaction ensemble} \\ \specialrule{.1em}{.1em}{.1em} 
\multicolumn{7}{c}{framework--OH $+$ H$_2$O $\rightleftharpoons$ framework--O$^-$ $+$ H$_3$O$^+$} \\
\rowcolor[HTML]{EFEFEF} 
\multicolumn{2}{c}{\cellcolor[HTML]{EFEFEF}quantity} & \multicolumn{4}{c}{\cellcolor[HTML]{EFEFEF}$\Delta\mu_{\text{rxn}}$ {[}kJ/mol{]}} & refs \\
\multicolumn{2}{c}{reaction chemical potential$^{b}$} & \multicolumn{4}{c}{$-700$ to $-300$} & this work \\
\end{tabular}
\caption{\textbf{Parameters of water, hydronium and the Brønsted acid zeolite.}
$^{a}$Lennard--Jones parameters for Al are taken from Si in TraPPE--Zeo
\cite{bai2013trappe}; its partial charge is lowered by one elementary charge
relative to Si to balance the compensating proton added at the acid site, so
that the framework remains neutral overall.
$^{b}$$\Delta\mu_{\text{rxn}}$ is the reaction chemical potential difference of
Eq.~\ref{rxmc_eqn} and is the only adjustable parameter of the reaction. It was
applied as $\mu(\mathrm{H_3O^+})$ with the remaining three species set to zero,
which is equivalent since only the combination enters the acceptance criterion.
The range quoted is the sweep used to locate the midpoint $\mu^{*}$
(Section~\ref{mu_sweep}) with a mixture
of equal reactants and products.}
\label{FFparamsChap3}
\end{table*}

\section*{Acknowledgements}
This material is based on work supported by the National Science Foundation under Grant \#2138938 for an internship at the National Institute of Standards and Technology (NIST). The material is also based upon work supported by, or in part by, the Army Research Laboratory and the Army Research Office under contract/grant number W911NF-24-1-0399.
This work was supported through a cooperative research and develop agreement between NIST and the University of Maryland Baltimore County.
NIST authors were solely funded by the United States of America government.
Certain commercial firms and trade names are identified in order to specify the usage procedures adequately for reproducibility.
Such identification is not intended to imply recommendation or endorsement by NIST, nor is it intended to imply that related products are necessarily the best available for the purpose.

\section*{Conflicts of Interest}
The authors declare that they have no known competing financial interests or personal relationships that could have appeared to influence the work reported in this paper.

\bibliography{references} 

\end{document}